\documentclass[onecolumn,showpacs,10pt]{revtex4}

\usepackage{graphicx}
\usepackage{dcolumn}
\usepackage{bm}

\input epsf

\begin{document}

\title{\Large Accretions of Various Types of Dark Energies onto Morris-Thorne Wormhole}

\author{\bf Ujjal Debnath\footnote{ujjaldebnath@yahoo.com ,
ujjal@iucaa.ernet.in}}

\affiliation{Department of Mathematics, Bengal Engineering and
Science University, Shibpur, Howrah-711 103, India.\\}

\date{\today}

\begin{abstract}
In this work, we have studied accretion of the dark energies onto
Morris-Thorne wormhole. For quintessence like dark energy, the
mass of the wormhole decreases and phantom like dark energy, the
mass of wormhole increases. We have assumed two types of dark
energy like variable modified Chaplygin gas (VMCG) and generalized
cosmic Chaplygin gas (GCCG). We have found the expression of
wormhole mass in both cases. We have found the mass of the
wormhole at late universe and this is finite. For our choices the
parameters and the function $B(a)$, these models generate only
quintessence dark energy (not phantom) and so wormhole mass
decreases during evolution of the universe. Next we have assumed 5
kinds of parametrizations of well known dark energy models. These
models generate both quintessence and phantom scenarios. So if
these dark energies accrete onto the wormhole, then for
quintessence stage, wormhole mass decreases upto a certain value
(finite value) and then again increases to infinite value for
phantom stage during whole evolution of the universe. We also
shown these results graphically.
\end{abstract}


\maketitle

\section{Introduction}

In recent observations it is strongly believed that the universe
is experiencing an accelerated expansion. The type Ia Supernovae
and Cosmic Microwave Background (CMB) \cite{Perlmutter,Riess}
observations have shown the evidences to support cosmic
acceleration. This acceleration is caused by some unknown matter
which has the property that positive energy density and negative
pressure satisfying $\rho+3p<0$ is dubbed as ``dark energy'' (DE)
\cite{Briddle,Spergel,Peebles,Cald}. If $\rho+p<0$, it is dubbed
as ``phantom energy''. The combined astrophysical observations
suggests that universe is spatially flat and the dark energy
occupies about 70\% of the total energy of the universe, the
contribution of dark matter is $\sim$ 26\%, the baryon is 4\% and
negligible radiation. A cosmological property in which there is an
infinite expansion in scale factor in a finite time termed as `Big
Rip'. In the phantom cosmology, big rip is a kind of future
singularity in which the energy density of phantom energy will
become infinite in a finite time. To realize the Big Rip scenario
the condition $\rho+p<0$ alone is not sufficient \cite{McInnes}.
Distinct data on supernovas showed that the presence of phantom
energy with $-1.2 < w < -1$ in the Universe is highly likely
\cite{Alam2}. In this case the cosmological phantom energy density
grows at large times and disrupts finally all bounded objects up
to subnuclear scale.\\

A wormhole is a feature of space that is essentially a
``shortcut'' from one point in the universe to another point in
the universe, allowing travel between them that is faster than it
would take light to make the journey through normal space. So the
wormholes are tunnels in spacetime geometry that connect two or
more regions of the same spacetime or two different spacetimes
\cite{Morris}. Wormholes may be classified into two categories -
Euclidean wormholes  and Lorentzian wormholes. The Euclidean
wormholes arise in Euclidean quantum gravity and the Lorentzian
wormholes \cite{thorne} which are static spherically symmetric
solutions of Einstein's general relativistic field equations
\cite{Visser1}. In order to support such exotic wormhole
geometries, the matter violating the energy conditions (null, weak
and strong), but average null energy condition is satisfied in
wormhole geometries \cite{visser,w1,w2}. For small intervals of
time, the weak energy condition (WEC) can be satisfied
\cite{viss}. Also the traversable wormhole solutions of the field
equations are obtained \cite{fer}. Recently evolving wormhole solutions
and their implications are discussed by several authors
\cite{kar,Anchordoqui,lobo,Cataldo,cat,cat2,Farooq,deb}.\\

The equations of motion for steady-state spherical symmetric flow
of matter into or out of a condensed object (e.g. neutron stars,
`black holes', etc.) are discussed by Michel \cite{Michel} and
also obtained analytic relativistic accretion solution onto the
static Schwarzschild black hole. The accretion of phantom dark
energy onto a Schwarzschild black hole was first modelled by
Babichev et al \cite{Babichev,Babichev1}. They established that
black hole mass will gradually decrease due to strong negative
pressure of phantom energy and finally all the masses tend to zero
near the big rip where it will disappear. Accretion of phantom
like variable modified Chaplygin gas onto Schwarzschild black hole
was studied by Jamil \cite{Jamil} who showed that mass of the
black hole will decrease when accreting fluid violates the
dominant energy condition and otherwise will increase. Also the
accretion of dark energy onto the more general Kerr-Newman black
hole was studied by Madrid et al \cite{Pedro} and Bhadra et al
\cite{Bhadra}. Till now, several authors
\cite{Chak,Maj,Nayak,Dwiv,Lima,Sharif,Sharif1,Sun,Kim,Mar,Sharif2,Rod,Abhas,Abhas1,Mar1}
have discussed the accretion of various components of dark energy
onto black holes. Recently, there is a great interest of the study
of dark energy accretion onto static wormhole
\cite{Diaz1,Far,Diaz2}. The phantom energy accretion onto wormhole
is discussed by Gonz$\acute{a}$lez-D$\acute{i}$az \cite{Diaz3}.
Madrid et al \cite{Mad} studied the dark energy accretion onto
black holes and worm holes phenomena could lead to unexpected
consequences, allowing even the avoidance of the considered
singularities. Also Mart$\acute{i}$n-Moruno \cite{Mor} have
considered a general formalism for the accretion of dark energy
onto astronomical objects, black holes and wormholes. It has been
shown that in models with four dimensions or more, any singularity
with a divergence in the Hubble parameter may be avoided by a big
trip, if it is assumed that there is no coupling between the bulk
and this accreting object. If this is not the case in more than
four dimensions, the evolution of the cosmological object depends
on the particular model. The dark energy accretion onto wormhole
in accelerating universe has also been discussed
\cite{Diaz4,Diaz5} recently.\\

In the following section, we assume the Morris-Thorne static
wormhole in presence of dark energy filled universe. If dark
energy accretes onto the wormhole, the rate of change of mass of
the wormhole is expressed in terms of the density and pressure of
dark energy and also find the expression of wormhole mass in terms
of density. Our main motivation of the work is to examine the
natures of the mass of the wormhole during expansion of the
universe if several kinds of dark energies accrete around the
wormhole. The candidates of dark energy are assumed to be variable
modified Chaplygin gas (VMCG) and generalized cosmic Chaplygin gas
(GCCG). Also we have assumed some kinds of parametrizations of
dark energy candidates. The mass of the wormhole has been
calculated for all types of dark energies and its natures have
been analyzed during expansion of the universe. Finally, we give
some concluding remarks of the whole work.\\

\section{Accretion Phenomena of Dark Energy onto Wormhole}

Let us consider a spherically symmetrical accretion of the dark
energy onto the wormhole. We consider a non-static spherically
symmetric Morris-Thorne wormhole metric \cite{Morris} given by
\begin{eqnarray}
ds^2=-e^{\Phi(r)}dt^{2}+\frac{dr^{2}}{1-\frac{K(r)}{r}}
+r^2\left(d\theta^2+\sin^2 \theta d\phi^2\right)
\end{eqnarray}
where the functions $K(r)$ and $\Phi(r)$ are the shape
function and redshift function respectively of radial
co-ordinate $r$. If $K(r_{0})=r_{0}$, the radius $r_{0}$
is called wormhole throat radius. So we want to consider the
outward region such that $r_{0}\le r<\infty$.\\

A proper dark-energy accretion model for wormholes should be
obtained by generalizing the Michel theory \cite{Michel} to the
case of wormholes. Such a generalization has been already
performed by Babichev et al \cite{Babichev,Babichev1} for the case
of dark-energy accretion onto Schwarzschild black holes. We shall
follow now the procedure used by Babichev et al
\cite{Babichev,Babichev1}, adapting it to the case of a
Morris-Thorne wormhole \cite{Diaz4}. For this purpose, we consider
the energy-momentum tensor for the dark energy (DE) in the form of
perfect fluid having the EoS $p=p(\rho)$, is
\begin{eqnarray}
T_{\mu\nu}=(\rho+p)u_\mu u_\nu+p g_{\mu\nu}
\end{eqnarray}

where $\rho$ and $p$ are the energy density and pressure of the
dark energy respectively and $u^\mu=\frac{dx^\mu}{ds}$ is the
fluid 4-velocity satisfying $u^\mu u_\mu=-1$. We assume that the
in-falling dark energy fluid does not disturb the spherical
symmetry of the wormhole. Now we assume $\Phi(r)=0$. The
relativistic Bernoulli's equation after the time component of the
energy-momentum conservation law $T^{\mu\nu}_{;\nu}=0$ provide the
first integral of motion for static, spherically symmetric
accretion onto wormhole which yields \cite{Diaz4}
\begin{eqnarray}
M^{-2}r^{2}u(\rho+p)\left(1-\frac{K(r)}{r} \right)^{-1}
 \left(u^2+\frac{K(r)}{r}-1\right)^{\frac{1}{2}}=C_{1}
\end{eqnarray}

where $M$ is the exotic mass of the wormhole,
$u=\frac{dr}{ds}~(>0)$ is the radial component of the velocity
four vector and the integration constant $C_{1}$ has the dimension
of the energy density.\\

Moreover, the second integration of motion is obtained from the
projection of the conservation law for energy-momentum tensor onto
the fluid four-velocity, $u_\mu T^{\mu \nu}_{;\nu}=0$, which gives
\cite{Diaz4}
\begin{eqnarray}
M^{-2}r^{2}u\left(1-\frac{K(r)}{r} \right)^{-\frac{1}{2}}
\exp{\left[\int^{\rho}_{\rho_\infty}\frac{d
\rho}{\rho+p(\rho)}\right]}=C_{2}
\end{eqnarray}
where $C_{2}~(>0)$ is dimensionless integration constant,
$\rho_{\infty}$ is the dark energy density at infinity. Further
the value of the constant $C_{2}$ can be evaluated for different
dark energy models. Now we denote $c_{s}^{2}=dp/d\rho$, the square
speed of sound of the accreted fluid. Critical point can be
calculated by taking logarithmic differential of equations (3) and
(4), and put the multipliers of $dr/r$ and $du/u$ to zero. After
the calculation, we obtain $c_{s}^{2}(r_{*})=0$ and $u_{*}=0$ at
the critical point $r=r_{*}$. So for this accretion, the critical
point cannot be found. If we include $\Phi(r)$ for accretion
process, the critical point may be found \cite{Dok}.\\

The rate of change of mass $\dot{M}$ of the exotic wormhole is
computed by integrating the flux of the dark energy over the
entire two dimensional surface of the wormhole i.e.,
$\dot{M}=\oint T_{t}^{r} dS$, where $T_{t}^{r}$ represents the
radial component of the energy-momentum densities and
$dS=\sqrt{-g} d\theta d\phi=r^{2}\sin\theta d\theta d\phi$
\cite{Babichev,Landau} is the element of the wormhole surface.
Using the above equations we obtain the rate of change of mass as
\cite{Diaz4}
\begin{eqnarray}
\dot{M}=-4 \pi Q M^2\left(1-\frac{K(r)}{r}
\right)^{\frac{1}{2}}(\rho+p)
\end{eqnarray}

where $Q$ is a positive constant. For the relevant asymptotic
regime $r\rightarrow\infty$, the above equation reduces to
\begin{eqnarray}
\dot{M}=-4 \pi Q M^2(\rho+p)
\end{eqnarray}

We see that the rate for the wormhole exotic mass due to accretion
of dark energy becomes exactly the negative to the similar rate in
the case of a Schwarzschild black hole, asymptotically. Since the
Morris-Thorne wormhole is static, so the mass of the wormhole
depends on $r$ only. When some fluid accretes outside wormhole,
the mass function $M$ of the wormhole is considered as a dynamical
mass function and hence it should be a function of time also. So
$\dot{M}$ of the equation (6) is time dependent and the increasing
or decreasing of the wormhole mass $M$ sensitively depends on the
nature of the fluid which accretes upon the wormhole. If
$\rho+p<0$ i.e., for phantom dark energy accretion, the mass of
the wormhole increases but if $\rho+p>0$ i.e., for non-phantom
dark energy accretion, the mass of the wormhole decreases. In the
following, we shall assume different types of dark energy models
such as variable modified Chaplygin gas, generalized cosmic
Chaplygin gas and parametrizations of some kinds of well known
dark energy models. The natures of mass function of wormhole will
be analyzed for present and future stages of expansion of the
universe when the above types of dark energies are accreting upon
wormhole.

\subsection{Variable Modified Chaplygin Gas as dark energy model}

We consider the background spacetime is spatially flat represented
by the homogeneous and isotropic FRW model of the universe which
is given by
\begin{equation}
ds^{2}=-dt^{2}+a^{2}(t)\left[dr^{2}+r^{2}(d\theta^{2}+sin^{2}\theta
d\phi^{2}) \right]
\end{equation}

where $a(t)$ is the scale factor. We assume the universe is filled
with Variable Modified Chaplygin Gas (VMCG) and the EoS is
\cite{Debnath} given by
\begin{equation}
p=A\rho-\frac{B(a)}{\rho^{\alpha}} ~~~~\text{with}~~~~ 0\le \alpha
\le 1,~A ~\text{is ~constant}>0.
\end{equation}
Here $B(a)$ is the function of scale factor $a$ and for simplicity
we choose $B(a)=B_{0}a^{-n}$, where $B_{0}>0$ and $n>0$ are
constants. \\

The Einstein's equations for FRW universe are (choosing $8\pi
G=c=1$)
\begin{eqnarray}
H^2 = \frac{1}{3} \rho~,
\end{eqnarray}
\begin{eqnarray}
\dot{H}=-\frac{1}{2}\left(p + \rho \right)
\end{eqnarray}

Conservation equation satisfied by the dark energy model VMCG is
\begin{eqnarray}
\dot{\rho}+3H(\rho+p)=0
\end{eqnarray}

where $H=\frac{\dot{a}}{a}$ is the Hubble parameter. Using
equations (7) and (10), we have the solution of $\rho$ as
\begin{equation}
\rho=\left[\frac{3(1+\alpha)B_{0}}{\{3(1+\alpha)(1+A)-n\}}~\frac{1}{a^{n}}+\frac{C}{a^{3(1+A)(1+\alpha)}}
\right]^{\frac{1}{1+\alpha}}
\end{equation}

where $C>0$ is an arbitrary integration constant and
$3(1+A)(1+\alpha)>n$, for positivity of first term.\\

Using equations (6), (9) and (11), we have \cite{Mor,Mad}
\begin{eqnarray}
\dot{M}=\frac{4 \pi}{\sqrt{3}} Q M^2\rho^{-\frac{1}{2}}\dot{\rho}
\end{eqnarray}
which integrates to yield
\begin{eqnarray}
M=\frac{M_{0}}{1-\frac{8\pi
QM_{0}}{\sqrt{3}}\left(\sqrt{\rho}-\sqrt{\rho_{0}} \right) }
\end{eqnarray}
where, $M_{0}$ and $\rho_{0}$ are the present values of the
wormhole mass and density of the dark energy respectively. Using
eq. (9), we get the mass function in terms of the Hubble parameter
$H$ as in the form \cite{Mor,Mad}
\begin{eqnarray}
M=\frac{M_{0}}{1-8\pi QM_{0}\left(H-H_{0}\right) }
\end{eqnarray}
where $H_{0}$ is the present value of the Hubble parameter. In the
late stage of the universe i.e., $a$ is very large $(z\rightarrow
-1)$, the mass of the wormhole will be
\begin{eqnarray}
M=\frac{M_{0}}{1+\frac{8\pi QM_{0}}{\sqrt{3}}~\sqrt{\rho_{0}} }
\end{eqnarray}

If we put the solution $\rho$ from equation (12) in equation (14),
the mass of wormhole $M$ can be expressed in terms of scale factor
$a$ and then use the formula of redshift $z=\frac{1}{a}-1$, $M$
will be in terms of redshift $z$. Now $M$ vs $z$ is drawn in
figure 1. Our choice of the function $B(a)$, the VMCG gives only
the quintessence dark energy, not phantom dark energy. So mass of
the wormhole always decreases for our case. From figure, we see
that $M$ decreases with $z$ dereases. So mass of wormhole
decreases if the VMCG accretes onto the wormhole.

\begin{figure}
\includegraphics[height=2.0in]{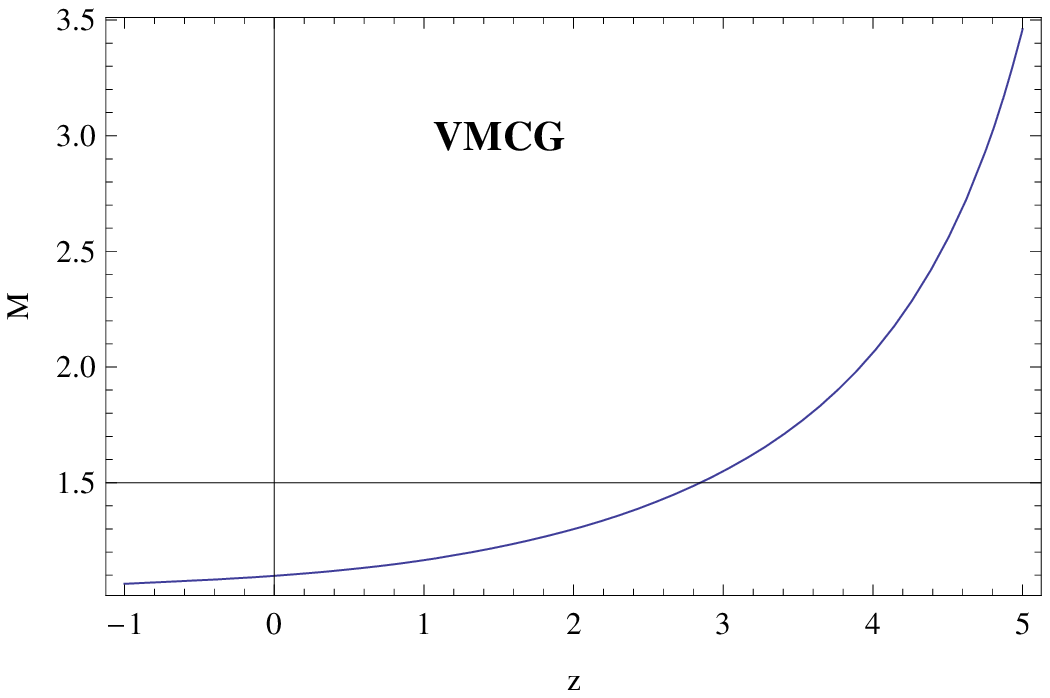}~~~~
\includegraphics[height=2.0in]{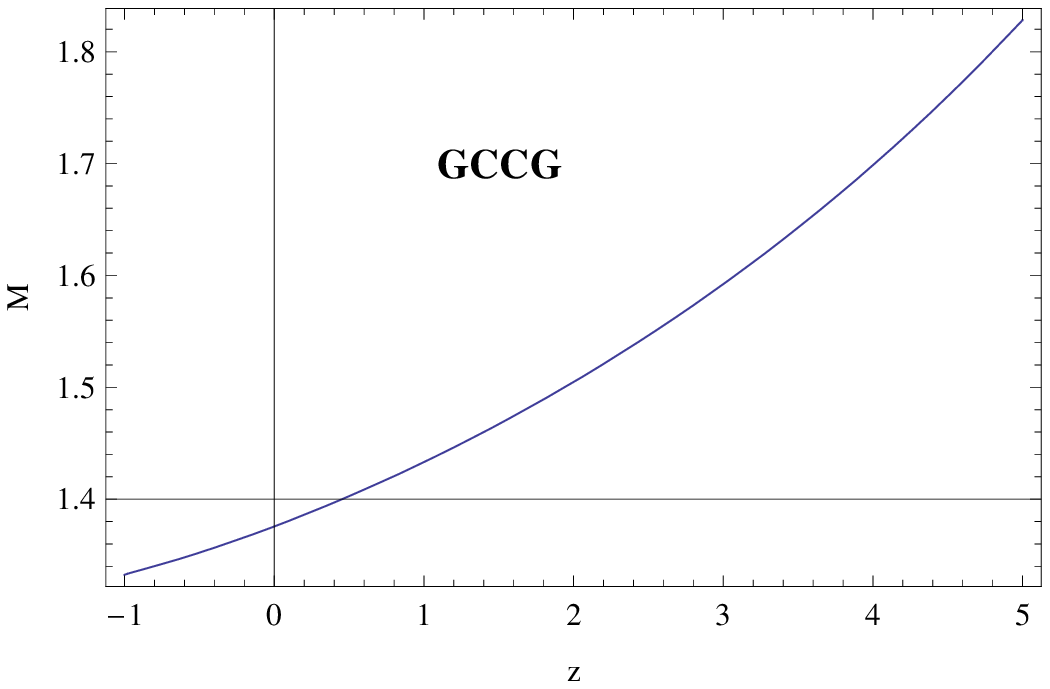}
\vspace{4mm}
~~~Fig.1~~~~~~~~~~~~~~~~~~~~~~~~~~~~~~~~~~~~~~~~~~~~~~~~~~~~~~~~~~~~~~~~~~Fig.2~~\\
\vspace{4mm} \vspace{.2in} Figs. 1 and 2 show the variations of
wormhole mass $M$ against redshift $z$ for VMCG and GCCG models.
\vspace{0.2in}
\end{figure}

\subsection{Generalized Cosmic Chaplygin Gas as the dark energy Model}

A new version of Chaplygin gas which is known as Generalized
Cosmic Chaplygin Gas (GCCG) \cite{Diaz0,Chak0} obeys the equation
of state
\begin{eqnarray}
p=-\rho^{-\alpha}\left[C+\left(\rho^{1+\alpha}-C\right)^{-w}\right]
\end{eqnarray}

where $C=\frac{A'}{(1+w)}-1$, $A'$ takes either positive or
negative constant, $-l<w<0$, $0\le\alpha\le 1$ and $l>1$. The EOS
reduces to that of current Chaplygin unified models for dark
matter and dark energy in the limit $w\rightarrow 0$ and satisfies
the conditions: (i) it becomes a de Sitter fluid at late time and
when $w=-1$, (ii) it reduces to $p=w \rho$ in the limit that the
Chaplygin parameter $A'\rightarrow 0$, (iii) it also reduces to
the EOS of current Chaplygin unified dark matter models at high
energy density and (iv) the evolution of density perturbations
derived from the chosen EOS becomes free from the pathological
behaviour of the matter power spectrum for physically reasonable
values of the involved parameters at late time. This EOS shows
dust era in the past and $\Lambda$CDM in the future.\\

From the conservation equation (10), we have the expression for
energy density of GCCG in the form \cite{Diaz0,Chak0}
\begin{eqnarray}
\rho=\left[C+\left(1+\frac{B}{a^{3(1+\alpha)(1+w)}}\right)^{\frac{1}{1+w}}\right]^{\frac{1}{1+\alpha}}
\end{eqnarray}
In the late stage of the universe i.e., $a$ is very large
$(z\rightarrow -1)$, the mass of the wormhole will be
\begin{eqnarray}
M=\frac{M_{0}}{1-\frac{8\pi
QM_{0}}{\sqrt{3}}\left((1+C)^{\frac{1}{2(1+\alpha)}}-\sqrt{\rho_{0}}
\right) }
\end{eqnarray}

If we put the solution $\rho$ from equation (18) in equation (14),
the mass of wormhole $M$ can be expressed in terms of scale factor
$a$ and hence $M$ will be in terms of redshift $z$. Now $M$ vs $z$
is drawn in figure 2. The GCCG gives only the quintessence dark
energy, not phantom dark energy. So mass of the wormhole always
decreases for our case. From figure, we see that $M$ decreases
with $z$ decreases. So mass of wormhole decreases if the GCCG
accretes onto the wormhole.

\subsection{Some Parameterizations of dark energy Models}

In astrophysical sense, the dark energy is popular to have a
redshift parametrization (i.e., taking the redshift $z$ as the
variable parameter of the EoS only) of the EoS as $p(z)=w(z)
\rho(z)$. The EoS parameter $w$ is currently constrained by the
distance measurements of the type Ia supernova and the current
observational data constrain the range of EoS as $-1.38<w<-0.82$
\cite{Melch}. Recently, the combination of WMAP3 and Supernova
Legacy Survey data shows a significant constraint on the EOS
$w=-0.97^{+0.07}_{-0.09}$ for the DE, in a flat universe
\cite{Sel}. Two mainstream families of red shift parametrizations
are considered here, viz.,\\

$(i)$ Family I: $w(z)=w_0 +w_1 \left(\frac{z}{1+z}\right)^n$~. In
this case, the conservation equation (11) gives the solution
\begin{equation}
\rho=\rho_{0}(1+z)^{3(1+w_{0})}e^{3(-1)^{-n}w_{1}\left\{Beta[1+z,-n,1+n]+\pi~
cosec ~n\pi \right\} }
\end{equation}

$(ii)$ Family II: $w(z)=w_0 +w_1 \frac{z}{\left(1+z\right)^n}$~.
The solution becomes
\begin{equation}
\rho=\rho_{0}(1+z)^{3(1+w_{0})}e^{-\frac{3w_{1}}{n(n-1)}\left[\frac{1+nz}{(1+z)^{n}}-1
\right]}
\end{equation}

where, $w_0$ and $w_1$ are two unknown parameters, which can be
constrained by the recent observations and $n$ is a natural
number. For different values of $n$, we will get following three
models of well known parametrizations (Models I, II, III). We
shall also assumed other two parametrizations (Models IV, V).
Since the following models generate both quintessence ($w(z)>-1$)
and phantom ($w(z)<-1$) dark energies for some suitable choices of
the parameters. So phantom divide is possible at $\Lambda$CDM
stage $w(z)=-1$. At the first stage, it occurs quintessence and
late stage it occurs phantom. So for quintessence stage, the mass
of the wormhole decreases and decreasing upto a certain limit (of
mass) and then again at phantom stage, the mass of the wormhole
increases. \\

$\bullet$ {\bf Model I (Linear):} For $n=0$, family II reduces to
the parametrization form $w(z)=w_{0}+w_{1}z$ \cite{Coor}. This is
known as $``Linear"$ parametrization. For Linear parametrization,
the solution becomes
\begin{equation}
\rho=\rho_{0}(1+z)^{3(1+w_{0}-w_{1})}e^{3w_{1}z}
\end{equation}
The above model generates phantom energy if $w(z)<-1$ i.e,
$z<-\frac{1+w_{0}}{w_{1}}$ provided $w_{1}>0$ and $w_{1}-w_{0}>1$.
If we drop this restriction, this model gives quintessence type
dark energy. Since this model is the phantom crossing model, so if
this dark energy accretes onto wormhole, for quintessence era,
wormhole mass decreases upto a certain limit and after that for
phantom era, the mass of the wormhole increases. We have shown
this scenario in figure 3. We see that wormhole mass $M$ decreases
for redshift $z$ decreases upto certain stage of $z$ ($\Lambda$CDM
stage) and then $M$ increases (phantom era) as universe expands.\\

$\bullet$ {\bf Model II (CPL):} For $n=1$, both the families I and
II lead to the same parametrization
$w(z)=w_{0}+w_{1}\frac{z}{1+z}$. This is known as $``CPL"$
parametrization \cite{Chev,Linder}. The solution becomes
\begin{equation}
\rho=\rho_{0}(1+z)^{3(1+w_{0}+w_{1})}e^{-\frac{3w_{1}z}{1+z}}
\end{equation}
The above model generates phantom energy if $w(z)<-1$ i.e,
$z<-\frac{1+w_{0}}{1+w_{1}}$ provided $w_{1}>-1$ and
$w_{1}-w_{0}>0$. If we drop this restriction, this model gives
quintessence type dark energy. This model is also the phantom
crossing model. From figure 4, we see that wormhole mass $M$
decreases for redshift $z$ decreases upto certain stage of $z$
($\Lambda$CDM stage) and then $M$ increases (phantom era) as
universe expands.\\

$\bullet$ {\bf Model III (JBP):} For family $II$, $n=2$ gives the
parametrization $w(z)=w_{0}+w_{1}\frac{z}{(1+z)^{2}}$. This is
known as $``JBP"$ parametrization \cite{Jassal}. The solution is
\begin{equation}
\rho=\rho_{0}(1+z)^{3(1+w_{0})}e^{\frac{3w_{1}z^{2}}{2(1+z)^{2}}}
\end{equation}
The above model generates phantom energy if $w(z)<-1$ i.e,
$z<-1+\frac{\sqrt{4(1+w_{0})w_{1}+w_{1}^{2} } }{2(1+w_{0})}$
provided $w_{0}>-1$ and $w_{1}<-4(1+w_{0})$. If we drop this
restriction, this model gives quintessence type dark energy. This
model is also the phantom crossing model. From figure 5, we see
that wormhole mass $M$ decreases for redshift $z$ decreases upto
certain stage of $z$ ($\Lambda$CDM stage) and then $M$ increases
(phantom era) as universe expands.\\

$\bullet$ {\bf Model IV:} Another type of parametrization is
considered as $ w(z)=-1+
\frac{A_{1}(1+z)+2A_{2}(1+z)^{2}}{3\left[A_{0}+A_{1}(1+z)+A_{2}(1+z)^{2}\right]}
$~~, where $A_{0},~A_{1}$ and $A_{2}$ are constants
\cite{Alam,Alam1}. This ansatz is exactly the cosmological
constant $w = -1$ for $A_{1} = A_{2} = 0$ and DE models with $w
=-2/3$ for $A_{0} = A_{2} = 0$ and $w = -1/3$ for $A_{0} = A_{1} =
0$. In this case, we get the solution
\begin{equation}
\rho=\frac{\rho_{0}[A_{0}+A_{1}(1+z)+A_{2}(1+z)^{2}]}{A_{0}+A_{1}+A_{2}}
\end{equation}
The above model generates phantom energy if $w(z)<-1$ i.e,
$z<-1-\frac{A_{1}}{A_{2}}$ provided $A_{0}<0$, $A_{1}>0$,
$A_{2}<0$ and $A_{0}+A_{1}+A_{2}<0$. For this condition, $\rho$ is
still positive. If we drop this restriction, this model gives
quintessence type dark energy. This model is also the phantom
crossing model. From figure 6, we see that wormhole mass $M$
decreases for redshift $z$ decreases upto certain stage
of $z$ and then $M$ increases (phantom era) as universe expands.\\

$\bullet$ {\bf Model V:} Other type of parametrization is assumed
to be $ w(z)=w_{0}+w_{1}log(1+z) $~\cite{Ef,Sil}. The solution is
obtained as
\begin{equation}
\rho=\rho_{0}(1+z)^{3(1+w_{0})}e^{\frac{3}{2}w_{1}[log(1+z)]^{2}}
\end{equation}
The above model generates phantom energy if $w(z)<-1$ i.e,
$z<-1+e^{-\frac{w_{0}}{w_{1}}}$ provided $w_{1}>0$. If we drop
this restriction, this model gives quintessence type dark energy.
This model is also the phantom crossing model. From figure 7, we
see that wormhole mass $M$ decreases for redshift $z$ decreases
upto certain stage of $z$ and then $M$ again increases (phantom
era) as universe expands.

\begin{figure}
\includegraphics[height=2.0in]{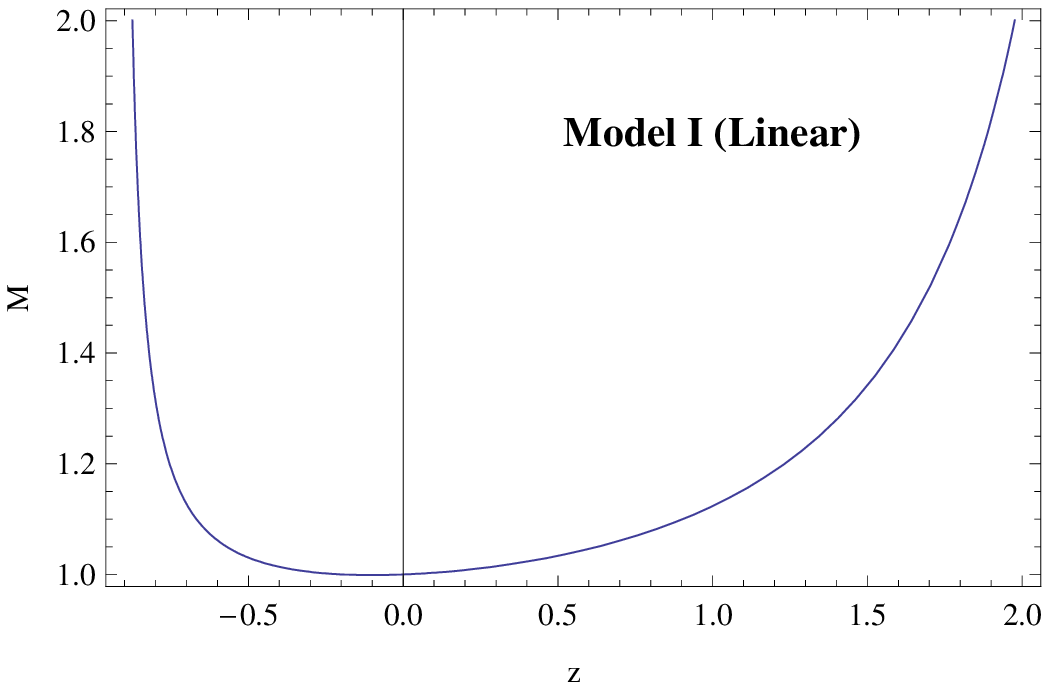}~~~~
\includegraphics[height=2.0in]{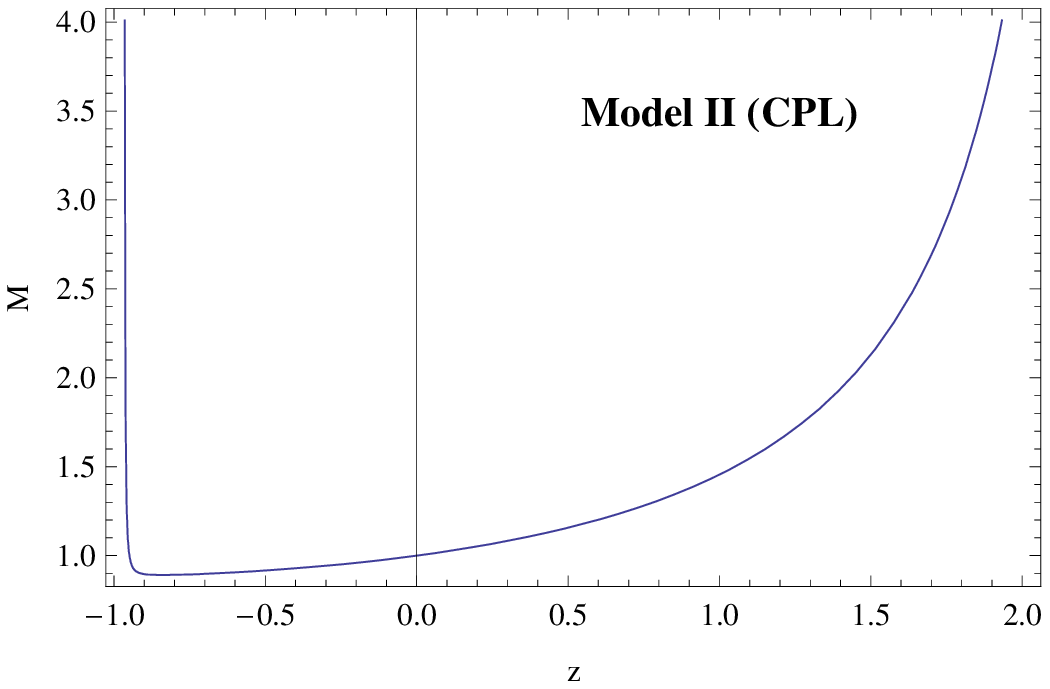}
\vspace{4mm}
~~~~~Fig.3~~~~~~~~~~~~~~~~~~~~~~~~~~~~~~~~~~~~~~~~~~~~~~~~~~~~~~~~~~~~~~~~~~~~~~Fig.4\\
\vspace{4mm}
\includegraphics[height=2.0in]{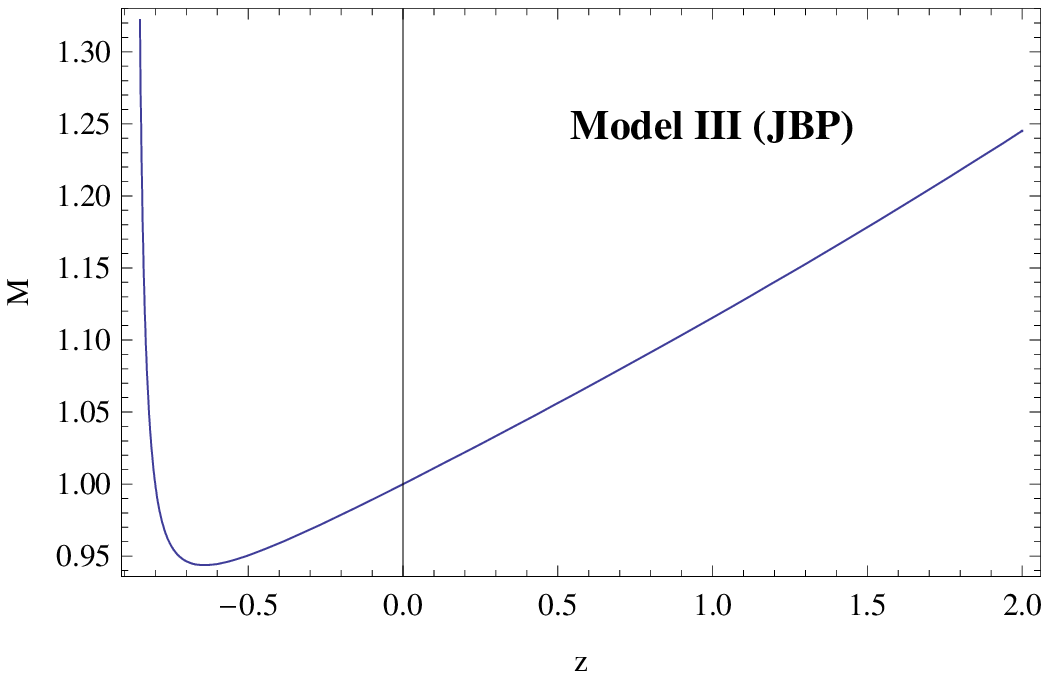}~~~~
\includegraphics[height=2.0in]{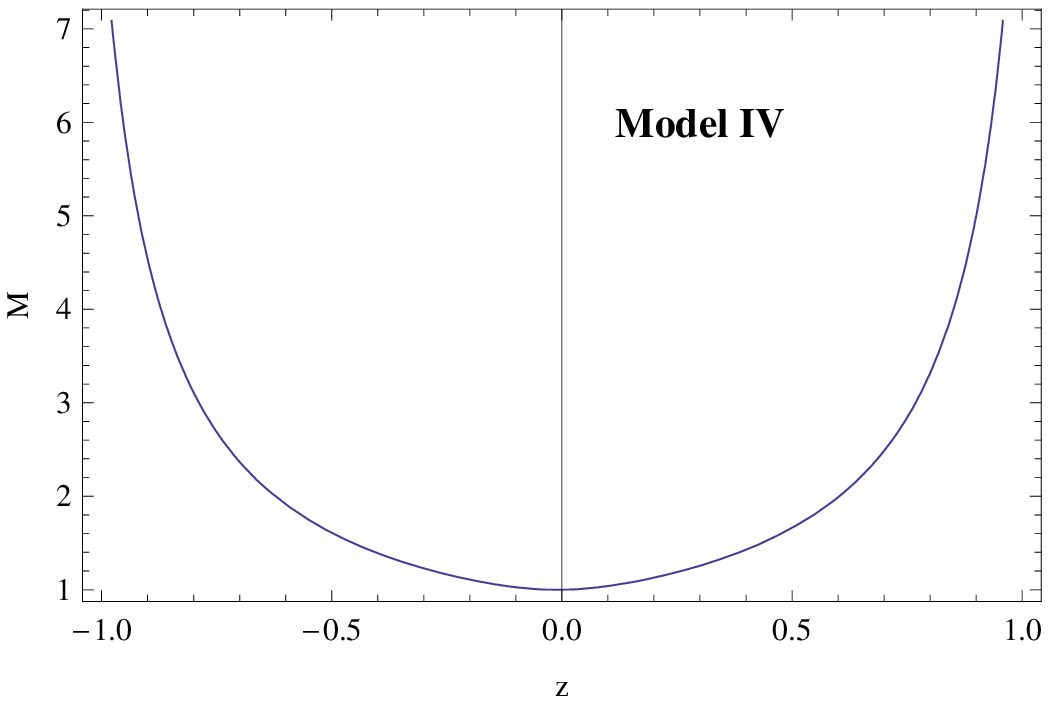}
\vspace{4mm}
~~~~~Fig.5~~~~~~~~~~~~~~~~~~~~~~~~~~~~~~~~~~~~~~~~~~~~~~~~~~~~~~~~~~~~~~~~~~~~~~Fig.6\\
\vspace{4mm}
\includegraphics[height=2.0in]{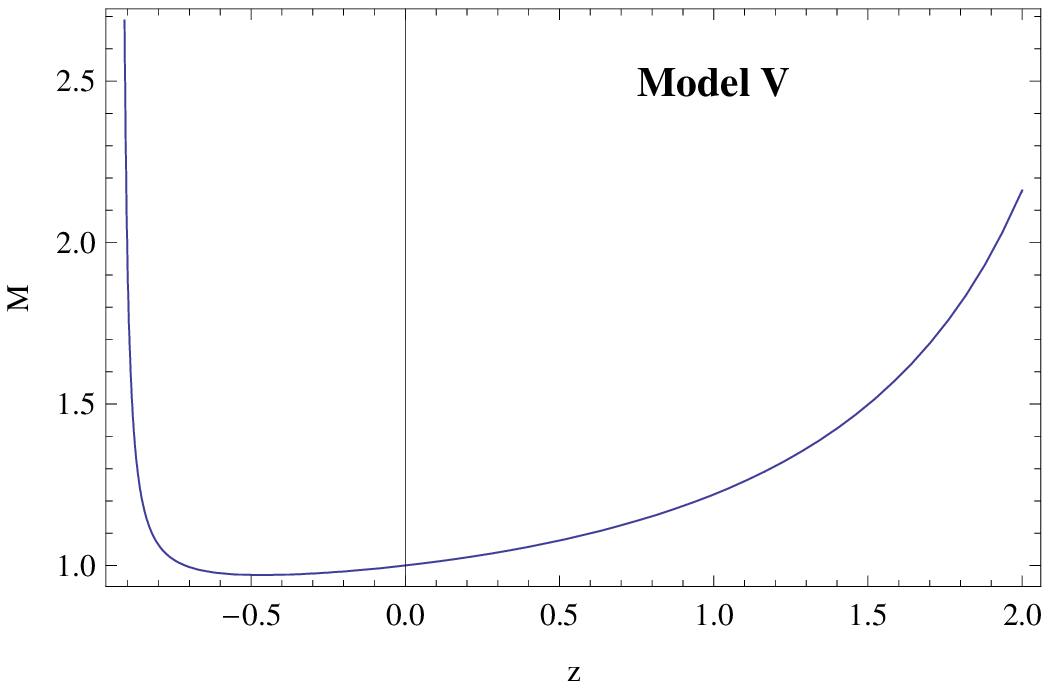}\\
\vspace{4mm} ~~~~~Fig.7 \vspace{4mm}

Figs. 3-7 show the variations of wormhole mass $M$ against
redshift $z$ for Models I-V respectively. \vspace{0.2in}
\end{figure}

\section{Discussions}

In this work, we have studied accretion of the dark energies onto
Morris-Thorne wormhole. A proper dark-energy accretion model for
wormholes have been obtained by generalizing the Michel theory
\cite{Michel} to the case of wormholes. Such a generalization has
been already performed by Babichev et al \cite{Babichev,Babichev1}
for the case of dark-energy accretion onto Schwarzschild black
holes. We have followed the procedure used by Babichev et al
\cite{Babichev,Babichev1}, adapting it to the case of a
Morris-Thorne wormhole. Here we have assumed the redshift function
$\Phi(r)=0$. We have shown that if $\Phi(r)=0$, the critical point
cannot be found. Astrophysically, mass of the wormhole is a
dynamical quantity, so the nature of the mass function is
important in our wormhole model for different dark energy filled
universe. The sign of time derivative of wormhole mass depends on
the signs of $\rho+p$. For quintessence like dark energy, the mass
of the wormhole decreases and phantom like dark energy, the mass
of wormhole increases. We have assumed recently proposed two types
of dark energy like variable modified Chaplygin gas (VMCG) and
generalized cosmic Chaplygin gas (GCCG). We have found the
expression of wormhole mass in both cases. We have found the mass
of the wormhole at late universe and this is found to be finite.
Our dark energy fluids violate the strong energy condition
($\rho+3p<0$ in late epoch), but do not violate the weak energy
condition ($\rho+p>0$). So the models drive only quintessence
scenario in late epoch, but do not generate the phantom epoch (in
our choice). So wormhole mass decreases during evolution of the
universe for these two dark energy models. Previously Babichev et
al \cite{Babichev} have shown that the mass of black hole
decreases due to phantom energy accretion. But for wormhole
accretion, the mass of wormhole increases due to phantom energy,
which is the opposite behaviour of black hole mass. Since our
considered dark energy candidates do not violate weak energy
condition, so the dynamical mass of the wormhole are decaying by
the accretion of our considered dark energies, though the
pressures of the dark energies are outside the wormhole. From
figures 1 and 2, we observe that the wormhole mass decreases as
$z$ increases for both VMCG and GCCG, which accrete onto the
wormhole in our expanding universe. Next we have assumed 5 kinds
of parametrizations (Models I-V) of well known dark energy models
(some of them are Linear, CPL, JBP models). These models generate
both quintessence and phantom scenarios for some restrictions of
the parameters. So if these dark energies accrete onto the
wormhole, then for quintessence stage, wormhole mass decreases
upto a certain value (finite value) and then again increases to
infinite value for phantom stage during whole evolution of the
universe. We also shown these results graphically clearly. Figures
3-7 show the mass of wormhole first decreases to finite value and
then increases to infinite value. In future work, it will be
interesting to show the natures of mass for various types of
wormhole models if different kinds of dark energies accrete upon
wormhole in accelerating universe also.\\

\section*{Acknowledgements}

The author is thankful to Institute of Theoretical Physics,
Chinese Academy of Science, Beijing, China for providing TWAS
Associateship Programme under which part of the work was carried
out. Also UD is thankful to CSIR, Govt. of India for providing
research project grant (No. 03(1206)/12/EMR-II). \\

\end{document}